\documentclass[12pt]{article}
\usepackage{epsfig}
\usepackage{cite}

\def\beq{\begin{equation}}
\def\eeq{\end{equation}}
\def\beqa{\begin{eqnarray}}
\def\eeqa{\end{eqnarray}}
 
\newlength{\dinwidth} \newlength{\dinmargin}
\begin{document}

\begin{center}
{\Large \bf Top-quark double-differential distributions at approximate N$^3$LO}
\end{center}
\vspace{2mm}
\begin{center}
{\large Nikolaos Kidonakis}\\
\vspace{2mm}
{\it Department of Physics, Kennesaw State University,\\
Kennesaw, GA 30144, USA}
\end{center}
 
\begin{abstract}
I present results for top-quark  double-differential distributions in transverse momentum and rapidity. Higher-order soft-gluon corrections from NNLL resummation are calculated through N$^3$LO. The corrections are large and they reduce theoretical uncertainties. The theoretical results are in good agreement with recent data from the LHC.

\end{abstract}
 
\section{Introduction}

Top-quark production is a central topic in particle physics due to the unique properties of the top quark and its potential role in exploring new physics. Both top-antitop production and single-top production have been intensively studied theoretically and experimentally for over two decades. Total cross sections, top-quark differential distributions in transverse momentum, rapidity, and invariant mass, as well as various asymmetries have all been calculated theoretically and measured experimentally at the Tevatron and the LHC to high precision (see Ref. \cite{NKtoprev} for a review). Top-quark double-differential distributions have also more recently been measured at the LHC \cite{CMSdiff13,CMS8tev,CMS13tev,ATLAS}; they provide yet another important test of the Standard Model. 

Top-quark differential distributions in $t{\bar t}$ production have been calculated at next-to-leading order (NLO) in Refs. \cite{NLOtt1,NLOtt2,NLOtt3}, and at next-to-next-to-leading order (NNLO) in Refs. \cite{NNLOttLHC,NNLOttTev,CDGKM}. Furthermore, soft-gluon corrections beyond the fixed-order results have been calculated for such distributions based on resummation at next-to-leading logarithm (NLL) accuracy in Refs. \cite{NKGS,NKnll} and at next-to-next-to-leading-logarithm (NNLL) accuracy in  Refs. \cite{AFNPY,NKnnll,AFNPY2,NKnnll2,NKn3lo}. 

The soft-gluon corrections are an important subset of the QCD corrections, and they are numerically dominant. These soft corrections in $t{\bar t}$ production provide excellent approximations at NLO and NNLO to the complete set of QCD corrections and, in fact, they predicted the NNLO results with very high accuracy - to the per mille level - for both the total cross section and the top-quark differential distributions in transverse momentum and rapidity. In going beyond the NNLO results, the soft-gluon corrections provide significant enhancements and a reduction of scale dependence when calculated at next-to-NNLO (N$^3$LO) \cite{NKn3lo,NKtt}. We denote results with these third-order soft-gluon corrections as approximate N$^3$LO (aN$^3$LO). Such aN$^3$LO results for top-quark single-differential distributions in transverse momentum and rapidity have appeared in \cite{NKn3lo}.

In this paper, we go beyond single-differential distributions and we provide new results for double-differential distributions in top-quark transverse momentum and rapidity through aN$^3$LO. The theoretical calculation of these distributions has now become important since there are recent relevant data from the LHC \cite{CMSdiff13,CMS8tev,CMS13tev,ATLAS}. In the next section, we provide a brief overview of the calculation. In Section 3, we provide results for the distributions and compare them with data from the LHC. We conclude in Section 4.

\section{Soft-gluon corrections}

We consider partonic processes in top-quark production, 
\beq
f_{1}(p_1)\, + \, f_{2}\, (p_2) \rightarrow t(p_t) \, + \, {\bar t}(p_{\bar t}) \, + \, X \, ,
\eeq
with $f_1$ and $f_2$ representing quarks or gluons in the colliding protons, and we define the usual partonic kinematical variables from the 4-momenta of the particles, 
$s=(p_1+p_2)^2$, $t=(p_1-p_t)^2$, and $u=(p_2-p_t)^2$.

We begin with a brief discussion of soft-gluon corrections for top-antitop pair production. 
These corrections come from the emission of soft (i.e. low-energy) gluons, and they arise from partial cancellations of infrared divergences between real-emission diagrams and diagrams with virtual quanta.
Soft-gluon corrections take the form of plus distributions of logarithms of 
a threshold variable which measures kinematical distance from partonic threshold. 
We note that partonic threshold is a generalized notion of threshold: the top-antitop pair is not necessarily produced at rest.

For $t{\bar t}$ production in single-particle-inclusive (1PI) kinematics, the partonic threshold variable is defined by $s_4=s+t+u-2m_t^2$, where  $m_t$ is the top-quark mass, and $s$, $t$, and $u$ are the partonic kinematical variables defined above. At partonic threshold, $s_4$ vanishes. For the $n$th-order corrections in the perturbative series, the soft-gluon terms involve plus distributions of the form 
$[(\ln^k(s_4/m_t^2))/s_4]_+$ with $0 \le k \le 2n-1$ at $n$th order in the strong coupling, $\alpha_s$.

The factorized form of the double-differential cross section in proton-proton collisions in 1PI kinematics is 
\beq
\frac{d^2\sigma^{pp \rightarrow t{\bar t}}}{dp_T \, dY}=\sum_{f_1,f_2} \; 
\int dx_1 \, dx_2 \,  \phi_{f_1/P_1}(x_1,\mu_F) \, 
\phi_{f_2/P_2}(x_2,\mu_F) \, 
{\hat \sigma}^{f_1 f_2 \rightarrow t{\bar t}}(s_4,s,t,u,\mu_F,\mu_R) \, ,
\eeq
where $p_T$ is the top-quark transverse momentum, $Y$ is the top-quark rapidity, $\mu_F$ and $\mu_R$ are the factorization and renormalization scales, respectively, $\phi_{f/P}$ are parton distribution functions (pdf) for parton $f$ in the proton, and ${\hat \sigma}^{f_1 f_2 \rightarrow t{\bar t}}$ is the hard-scattering partonic cross section. 

The resummation of soft-gluon corrections follows from the factorization of the cross section in moment space \cite{NKGS,LOS}. We define moments of the partonic cross section as 
${\hat\sigma}(N)=\int_0^{s_4^{\rm max}} (ds_4/s) \,  e^{-N s_4/s} \, {\hat\sigma}(s_4)$, and note that logarithms of $s_4$ transform in moment space into logarithms of $N$, with the latter exponentiating. We also define moments of the pdf as $\phi(N_i)=\int_0^1 e^{-N_i(1-x_i)} \phi(x_i) \, dx_i$. We also consider the parton-parton cross section $d^2\sigma^{f_1 f_2 \rightarrow t{\bar t}}/(dp_T \, dY)$ and define its moments as 
\beq
\frac{d^2\sigma^{f_1 f_2 \rightarrow t{\bar t}}(N)}{dp_T \, dY}=\int_0^{S_4^{\rm max}} 
\frac{dS_4}{S} \,  e^{-N S_4/S} \, \frac{d^2\sigma^{f_1 f_2 \rightarrow t{\bar t}}(S_4)}{dp_T \, dY} \, , 
\eeq
where $S_4/S=-(1-x_1)u/s-(1-x_2)t/s+s_4/s$. 
We then write the factorized moment-space parton-parton cross section in $4-\epsilon$ dimensions, as
\beq
\frac{d^2\sigma^{f_1 f_2 \rightarrow t{\bar t}}(N, \epsilon)}{dp_T \, dY}  
=\phi_{f_1/f_1}(N_1,\mu_F,\epsilon)\; 
\phi_{f_2/f_2}(N_2,\mu_F,\epsilon) \;
{\hat \sigma}^{f_1 f_2 \rightarrow t{\bar t}}(N,\mu_F,\mu_R) \, . 
\label{fac}
\eeq
 
A refactorized form of this cross section\cite{NKGS,LOS} is 
\beq
\frac{d^2\sigma^{f_1 f_2 \rightarrow t{\bar t}}(N, \epsilon)}{dp_T \, dY}= 
\left(\prod_{i=1,2}  J_i \left(N_i,\mu_F,\epsilon \right)\right) 
{\rm tr} \left\{H^{f_1 f_2\rightarrow t{\bar t}}\left(\alpha_s(\mu_R)\right) \, 
S^{f_1 f_2 \rightarrow t{\bar t}}\left(\frac{m_t}{N \mu_F},\alpha_s(\mu_R) \right)\right\} \, .
\label{refac}
\eeq
The infrared-safe hard function $H^{f_1 f_2\rightarrow t{\bar t}}$ does not depend on $N$, and it describes contributions 
from the amplitude and from the complex conjugate of the amplitude. 
The soft function $S^{f_1 f_2\rightarrow t{\bar t}}$ describes 
the emission of noncollinear soft gluons in the process. 
Both the hard and the soft functions are process-dependent matrices in color space in the partonic 
scattering. The $J_i$ denote functions that describe universal soft 
and collinear emission from the incoming partons.
 
Comparing Eqs. (\ref{fac}) and (\ref{refac}), we get the following expression for the moment-space hard-scattering partonic cross section,
\beqa
{\hat \sigma}^{f_1 f_2 \rightarrow t{\bar t}}(N,\mu_F,\mu_R)&=&
\frac{\prod_{i=1,2}  J_i \left(N_i,\mu_F,\epsilon \right)}{\phi_{f_1/f_1}(N_1,\mu_F,\epsilon)\; \phi_{f_2/f_2}(N_2,\mu_F,\epsilon)} 
\nonumber \\ && \times 
{\rm tr} \left\{H^{f_1 f_2\rightarrow t{\bar t}}\left(\alpha_s(\mu_R)\right) \, 
S^{f_1 f_2 \rightarrow t{\bar t}}\left(\frac{m_t}{N \mu_F},\alpha_s(\mu_R) \right)\right\} \, .
\label{sigN}
\eeqa

The $N$-dependence of the soft matrix $S^{f_1 f_2\rightarrow t{\bar t}}$ is resummed via renormalization group evolution\cite{NKGS}, 
\beq
S_b^{f_1 f_2\rightarrow t{\bar t}}=(Z_S^{f_1 f_2\rightarrow t{\bar t}})^{\dagger} \; S^{f_1 f_2\rightarrow t{\bar t}} \; Z_S^{f_1 f_2\rightarrow t{\bar t}}
\eeq
where $S_b^{f_1 f_2\rightarrow t{\bar t}}$ is the unrenormalized quantity 
and $Z_S^{f_1 f_2\rightarrow t{\bar t}}$ is a matrix of renormalization constants.
Thus, $S^{f_1 f_2\rightarrow t{\bar t}}$ obeys the renormalization group equation
\beq
\left(\mu \frac{\partial}{\partial \mu}
+\beta(g_s)\frac{\partial}{\partial g_s}\right) S^{f_1 f_2\rightarrow t{\bar t}}
=-(\Gamma_S^{f_1 f_2\rightarrow t{\bar t}})^{\dagger} \; S^{f_1 f_2\rightarrow t{\bar t}}-S^{f_1 f_2\rightarrow t{\bar t}} \; \Gamma_S^{f_1 f_2\rightarrow t{\bar t}}
\eeq
where $g_s^2=4\pi\alpha_s$ and $\beta$ is the QCD beta function.
The evolution of the soft function is controlled by the soft anomalous dimension matrix, $\Gamma_S^{f_1 f_2\rightarrow t{\bar t}}$,
which is calculated from the coefficients of the ultraviolet poles 
of eikonal diagrams.

The moment-space resummed cross section is derived  
from the renormalization-group evolution of the soft function and the other $N$-dependent functions in Eq. (\ref{sigN}), 
and it is given by \cite{NKtoprev,NKnll,NKnnll,NKGS,LOS}
\beqa
{\hat{\sigma}}^{f_1 f_2\rightarrow t{\bar t}}_{{\rm resum}}(N) &=&
\exp\left[ \sum_{i=1,2} E_{f_i}(N_i, \mu_F)\right] \,
{\rm tr} \left\{H^{f_1 f_2\rightarrow t{\bar t}}\left(\alpha_s(\sqrt{s})\right) \right.
\nonumber\\ && \hspace{-33mm} \left. \times \,
\exp \left[\int_{\sqrt{s}}^{{\sqrt{s}}/{\tilde N'}}
\frac{d\mu}{\mu} \; \Gamma_S^{\dagger \, f_1 f_2\rightarrow t{\bar t}}\left(\alpha_s(\mu)\right)\right] \; 
S^{f_1 f_2\rightarrow t{\bar t}} \left(\alpha_s\left(\frac{\sqrt{s}}{\tilde N'}\right)\right) \;
\exp \left[\int_{\sqrt{s}}^{{\sqrt{s}}/{\tilde N'}}
\frac{d\mu}{\mu}\; \Gamma_S^{f_1 f_2\rightarrow t{\bar t}}
\left(\alpha_s(\mu)\right)\right] \right\} \, .
\nonumber \\
\label{resummed}
\eeqa

The first exponential resums universal soft and collinear contributions from the incoming partons \cite{GS,CT}. The process-dependent hard and soft functions are known to one loop \cite{NKnll,AFNPY,AFNPY2}. The soft anomalous dimensions for the partonic processes $q{\bar q} \rightarrow t{\bar t}$ and $gg \rightarrow t{\bar t}$ are known at one \cite{NKGS} and two \cite{AFNPY,NKnnll,NK2loop,FNPY} loops.

We expand the NNLL resummed cross section to second and third orders and match to complete analytical NLO results. By doing an expansion, we avoid the problems with the divergence in the resummed expression which would require a prescription. As has been explained before \cite{NKnll}, such prescriptions have failed to accurately predict the correct size of the higher-order corrections, in contrast to our finite-order expansions. The double-differential distributions derived by adding second-order soft-gluon corrections to the complete NLO results are denoted as approximate NNLO (aNNLO); when third-order soft-gluon corrections are added to the aNNLO result, the distributions are denoted as aN$^3$LO. 
 
\section{Top-quark double-differential distributions}

We now provide theoretical predictions for the top-quark double-differential distributions in transverse momentum and rapidity through aN$^3$LO at LHC energies.

\begin{figure}[htbp]
\begin{center}
\includegraphics[width=83mm]{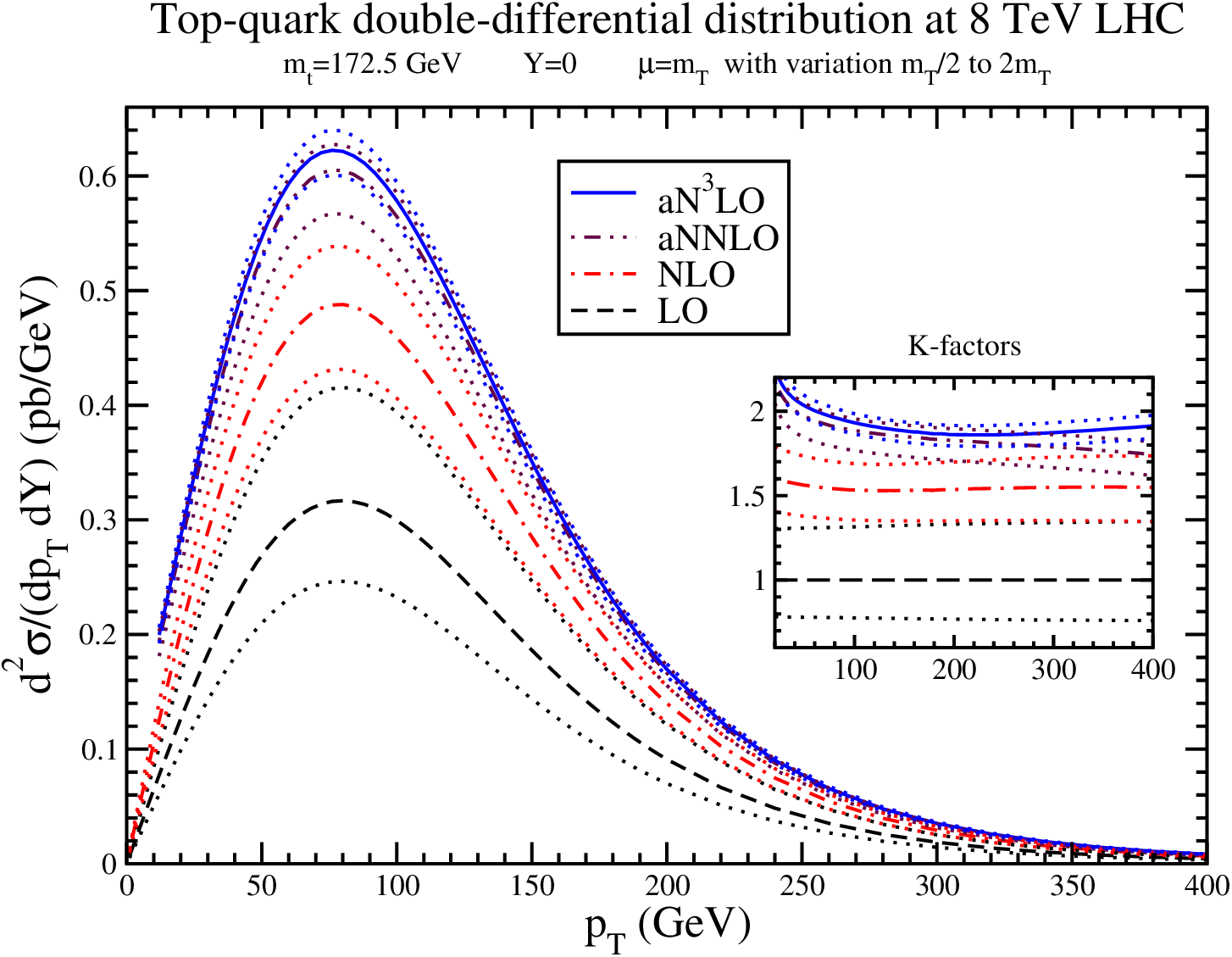}
\hspace{1mm}
\includegraphics[width=83mm]{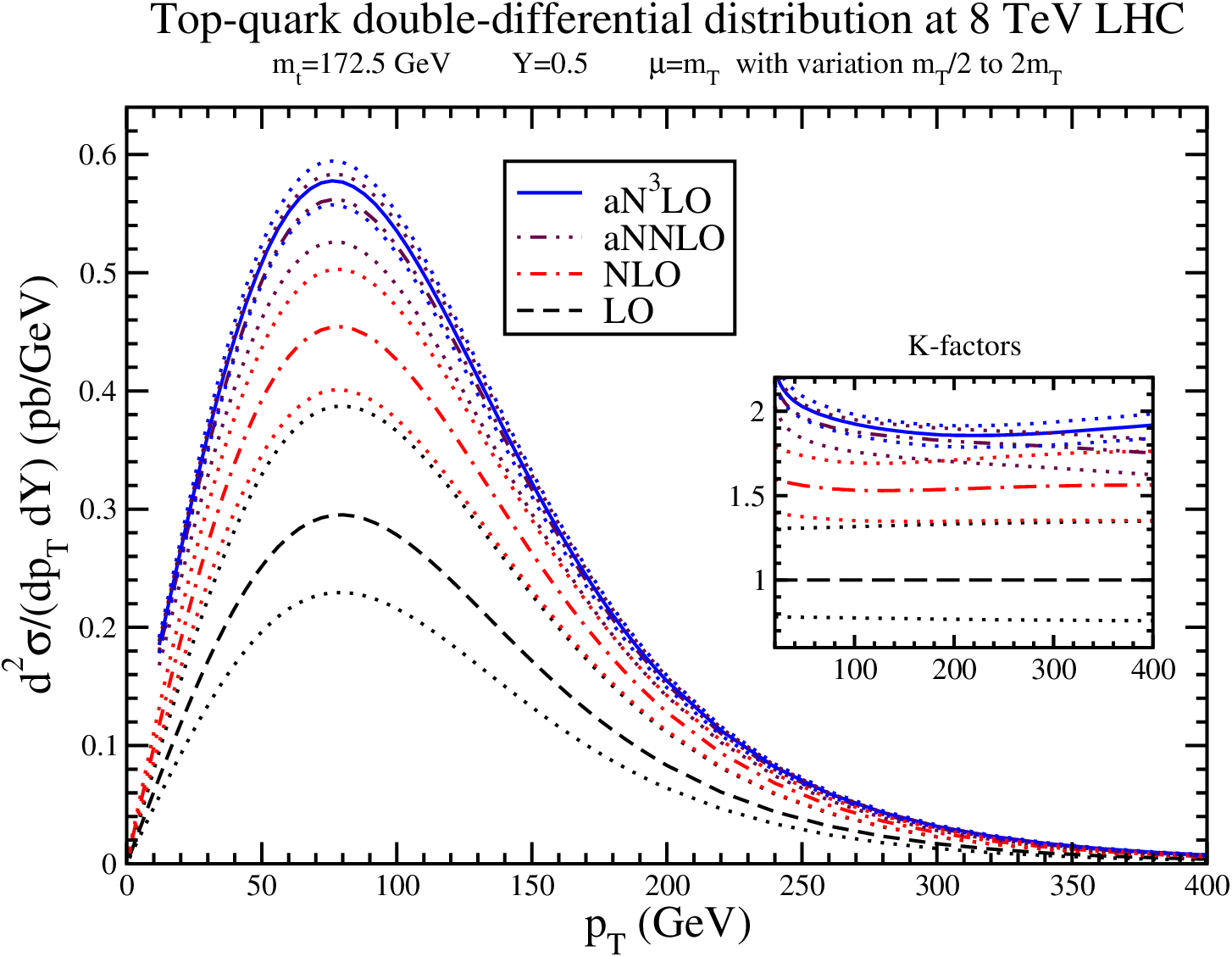}
\bigbreak
\includegraphics[width=83mm]{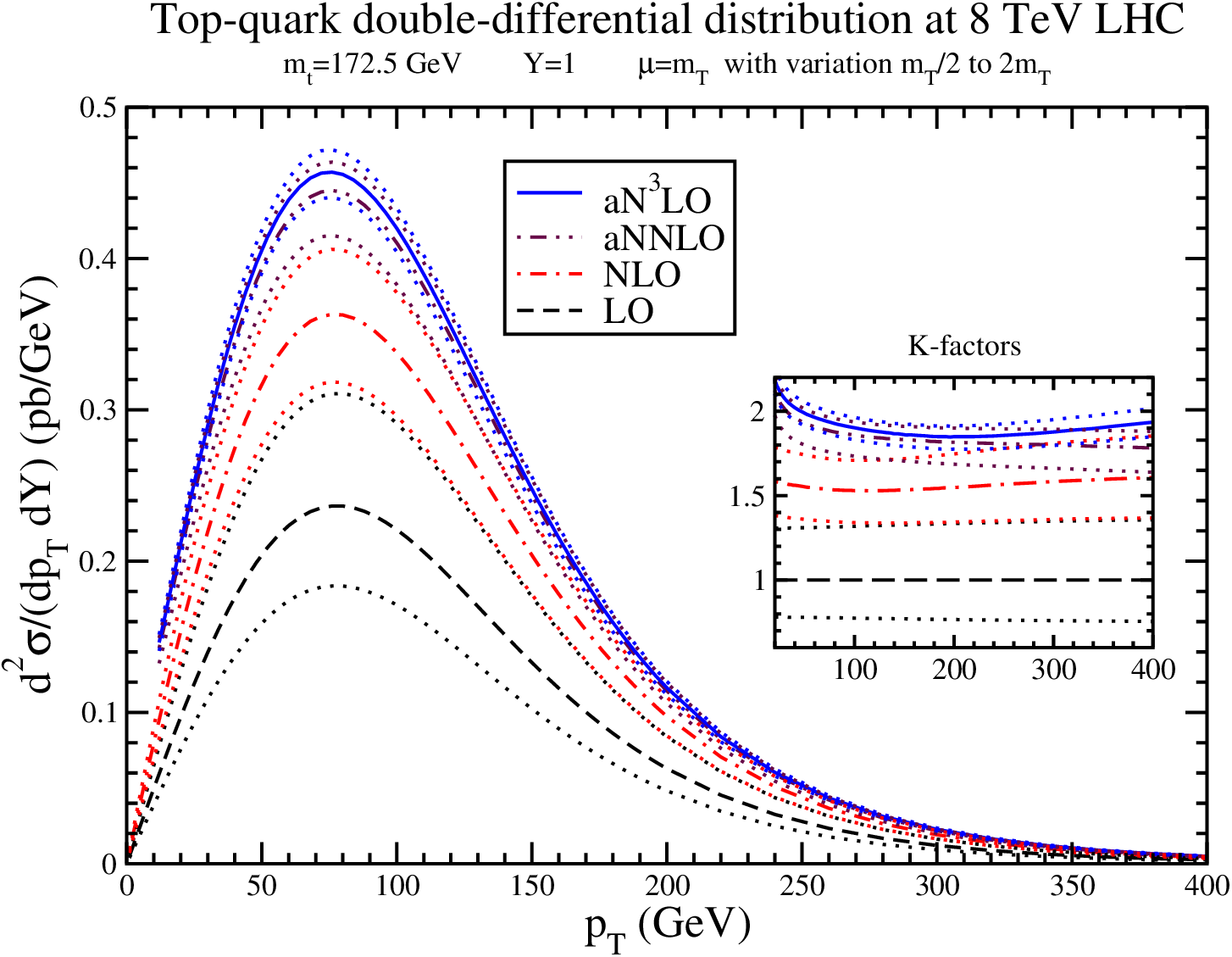}
\hspace{1mm}
\includegraphics[width=83mm]{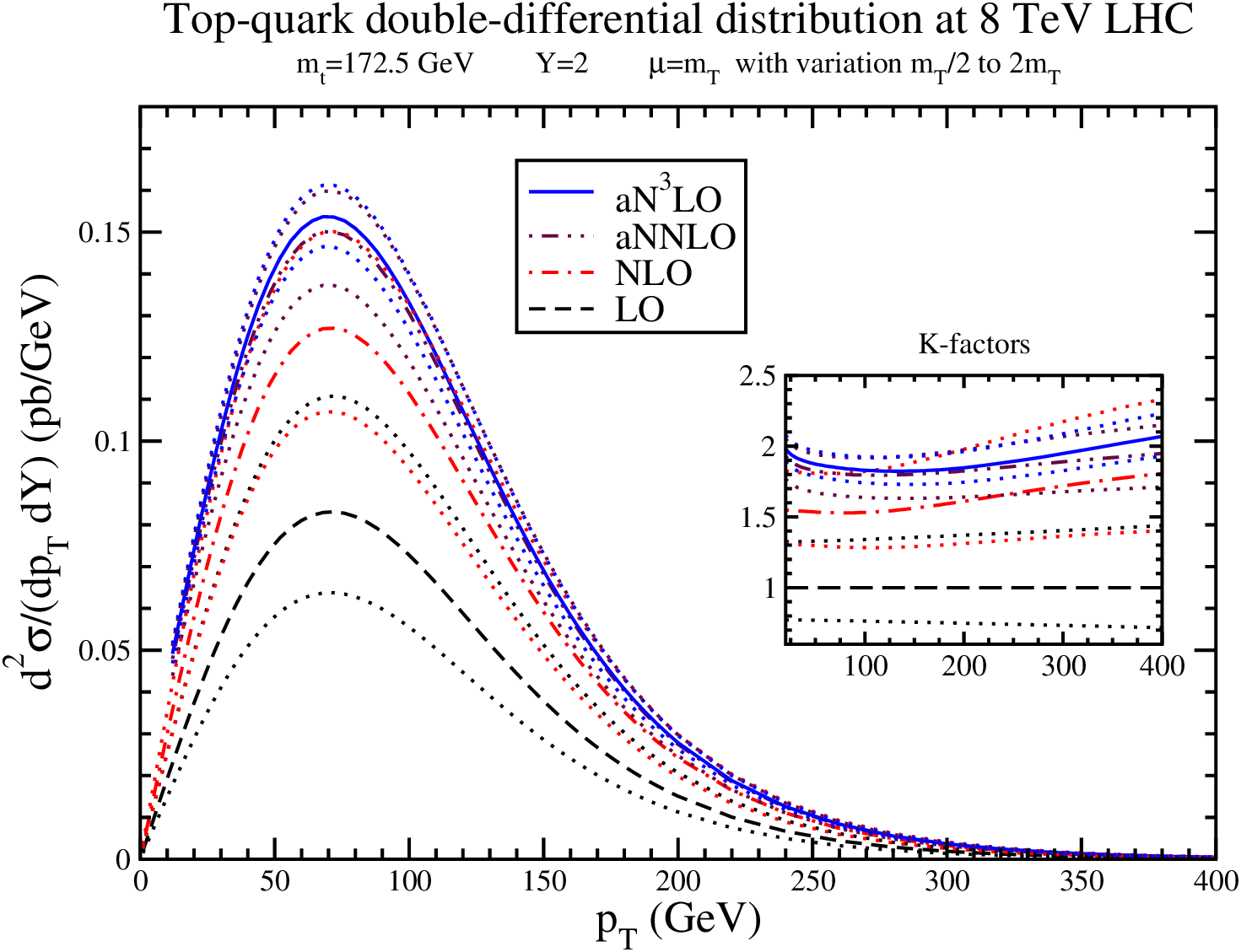}
\caption{The top-quark double-differential distributions, $d^2\sigma/(dp_T dY)$, at 8 TeV LHC energy, are displayed as functions of $p_T$ for four different values of rapidity. The LO, NLO, aNNLO, and aN$^3$LO results are shown with central scale $\mu=m_T$ (solid lines) and scale variation $m_T/2$ and $2m_T$ (dotted lines). The $K$-factors relative to LO are shown in the inset plots.}
\label{top8lhc}
\end{center}
\end{figure}

In Fig. \ref{top8lhc}, we show theoretical results for the double-differential distributions in top-quark transverse-momentum and rapidity, $d^2\sigma/(dp_T dY)$, at 8 TeV LHC energy as functions of the top-quark $p_T$ for four different values of the top-quark rapidity. The upper-left (upper-right) plot is for a top-quark rapidity of 0 (0.5) while the bottom-left (bottom-right) plot is for a top-quark rapidity of 1 (2). In each plot we show results at LO, NLO, aNNLO, and aN$^3$LO with a central scale choice of $\mu=m_T$, where $m_T=(p_T^2+m_t^2)^{1/2}$ is the transverse mass. We use MMHT2014 \cite{MMHT} NNLO pdf throughout, as we are interested in the growth of the perturbative series. The scale variation, $m_T/2 \le \mu \le 2 m_T$, at each order is shown by the dotted curves. We observe a reduction in scale variation as the order of the perturbative calculation is increased, as expected. 

The insets in the plots show the $K$ factors with respect to the central LO result. It is clear that the higher-order corrections are large at NLO and aNNLO, and still significant at aN$^3$LO, for all values of $p_T$ and rapidity. The aN$^3$LO/LO $K$-factor is around 2. The NLO corrections are naturally the largest, but there are significant contributions at aNNLO and even at aN$^3$LO. Again, it is clear from the inset plots that the scale uncertainty decreases significantly with each higher-order contribution.

\begin{figure}[htbp]
\begin{center}
\includegraphics[width=83mm]{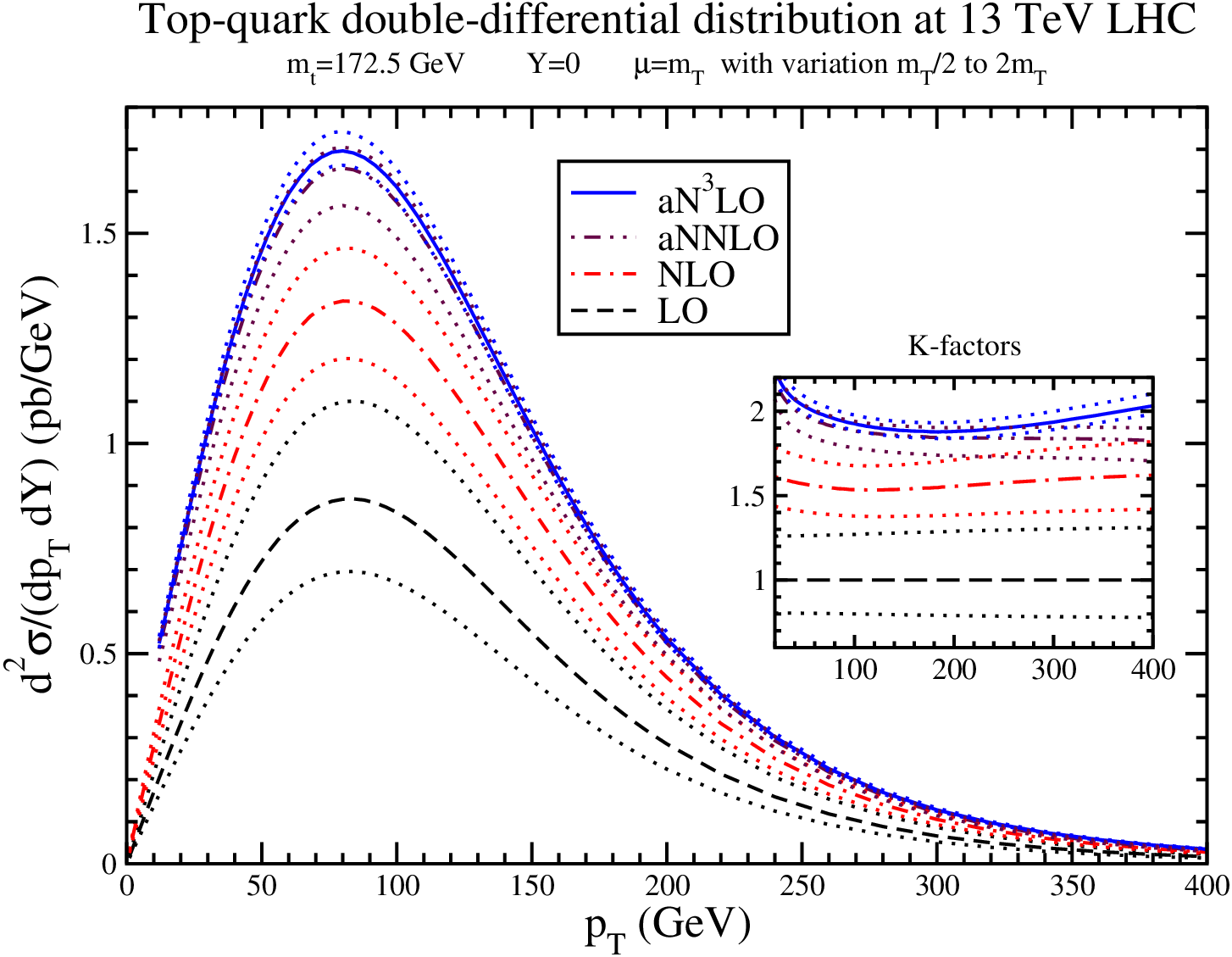}
\hspace{1mm}
\includegraphics[width=83mm]{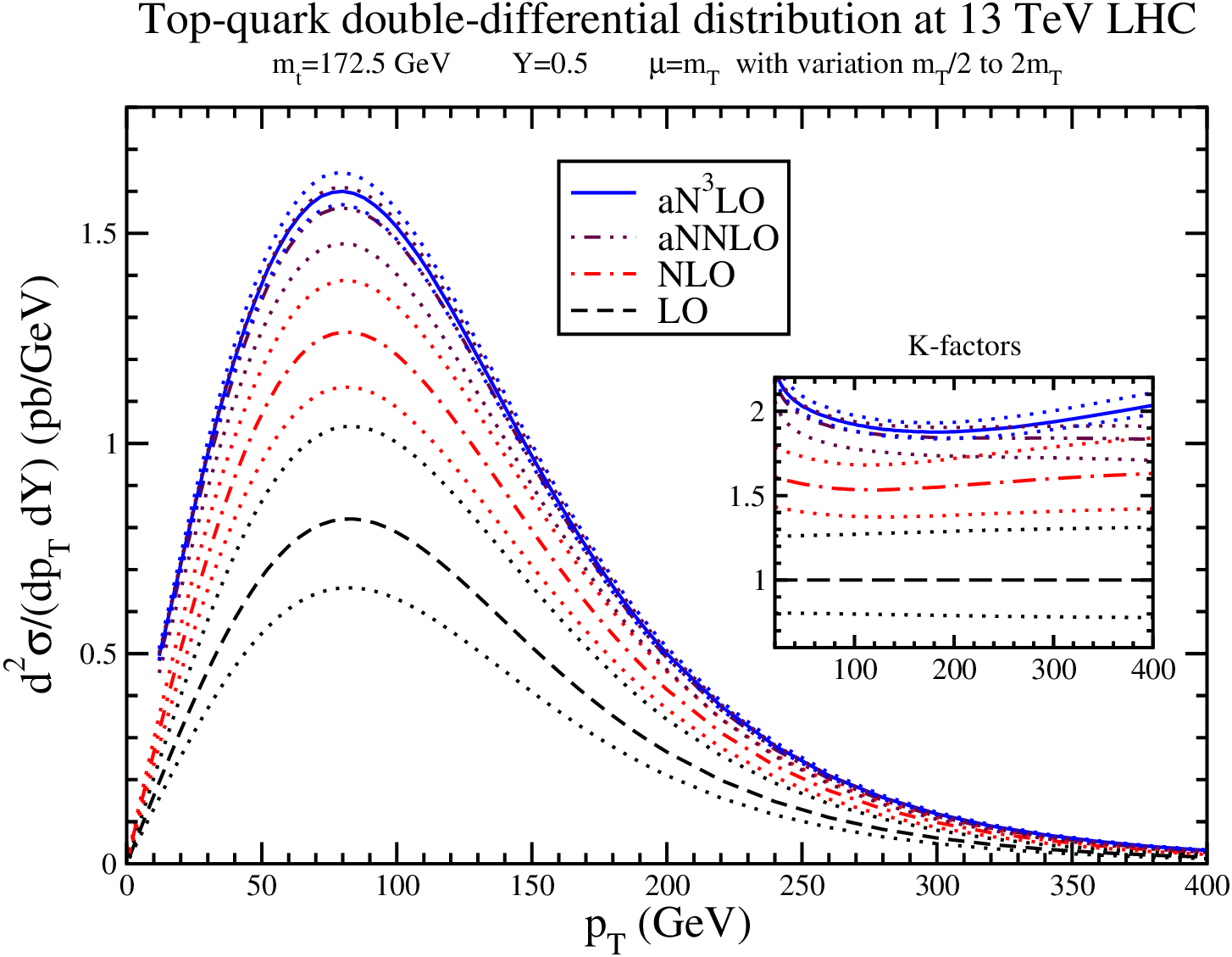}
\bigbreak
\includegraphics[width=83mm]{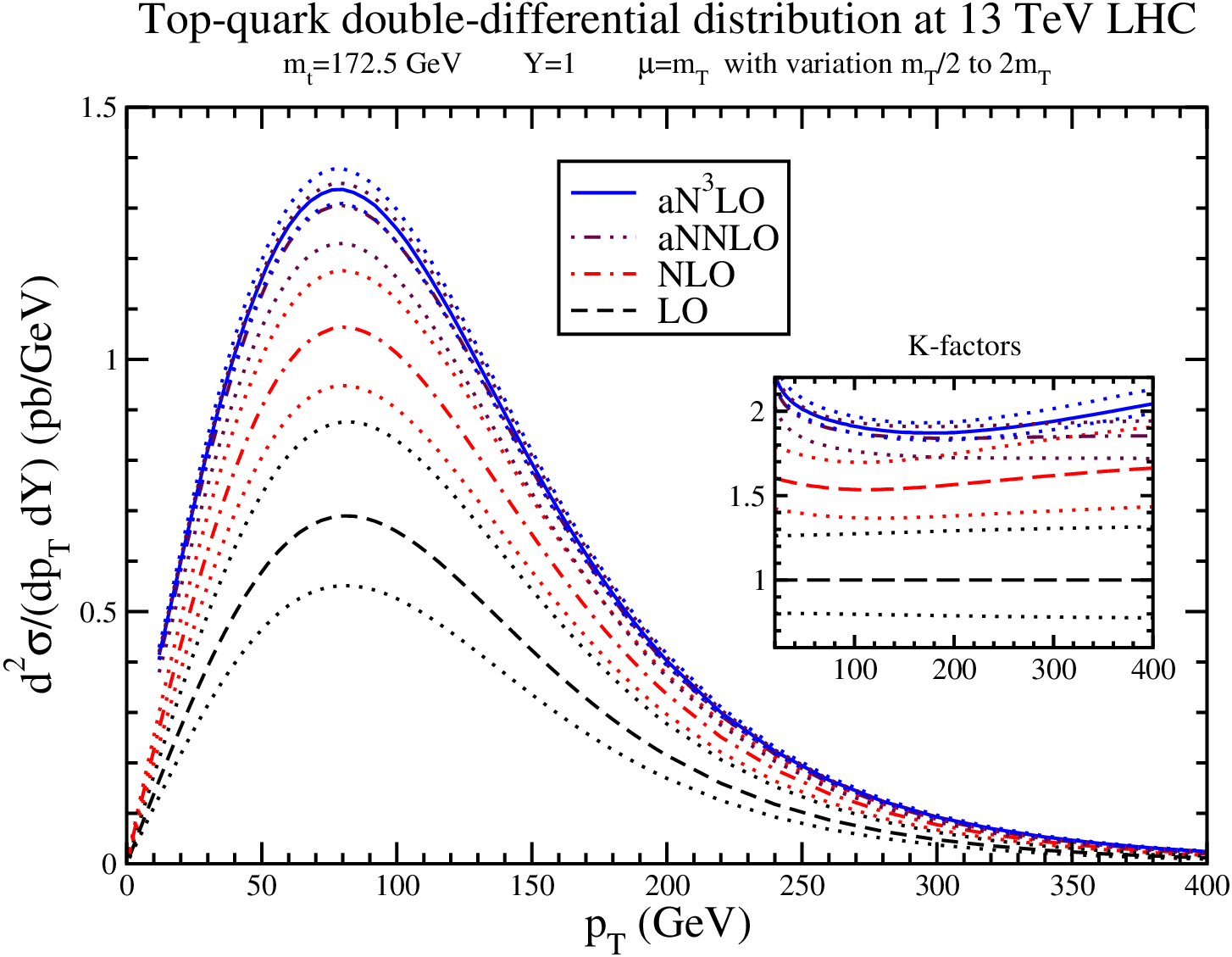}
\hspace{1mm}
\includegraphics[width=83mm]{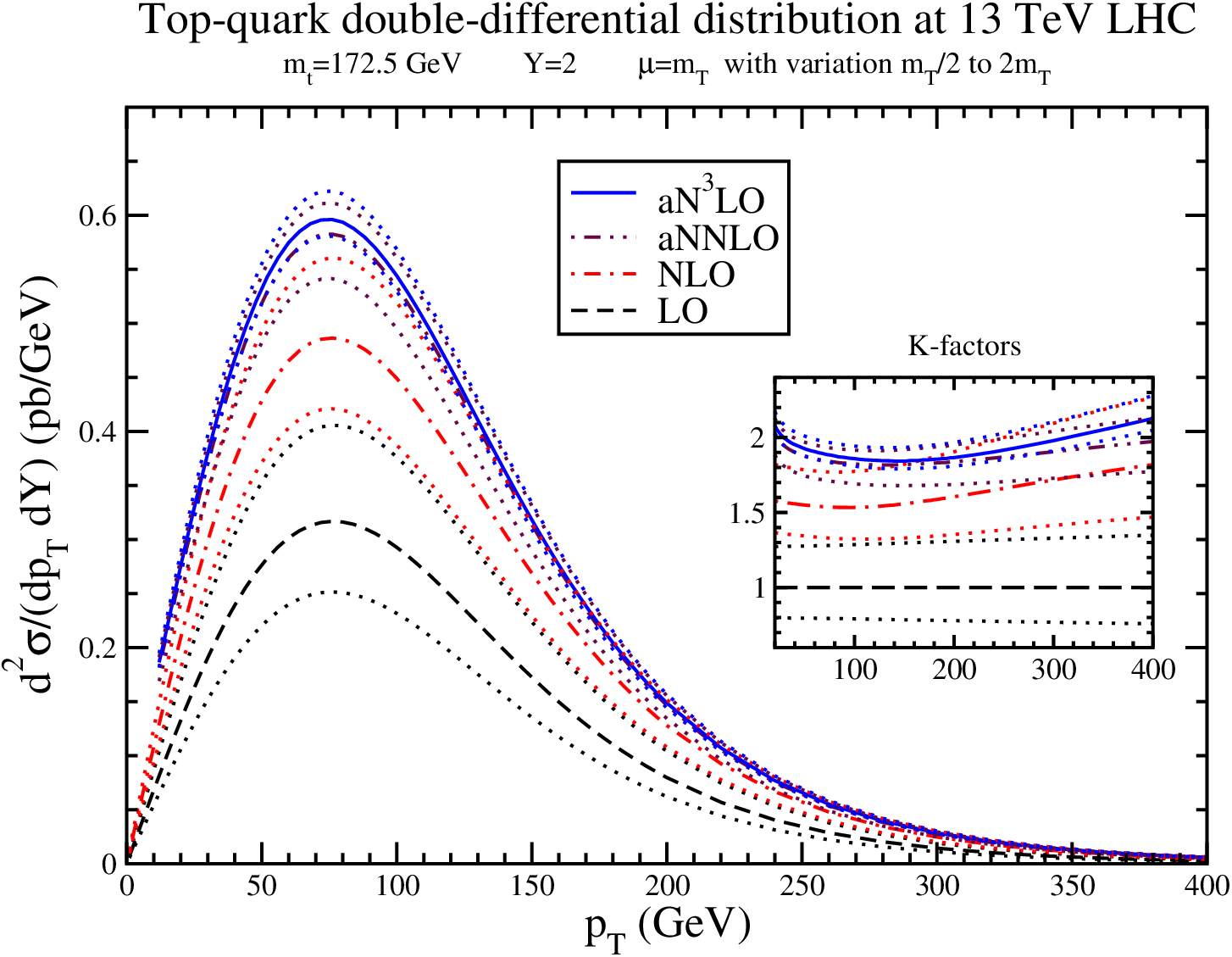}
\caption{The top-quark double-differential distributions, $d^2\sigma/(dp_T dY)$, at 13 TeV LHC energy, are displayed as functions of $p_T$ for four different values of rapidity. The LO, NLO, aNNLO, and aN$^3$LO results are shown with central scale $\mu=m_T$ (solid lines) and scale variation $m_T/2$ and $2m_T$ (dotted lines). The $K$-factors relative to LO are shown in the inset plots.}
\label{top13lhc}
\end{center}
\end{figure}

In Fig. \ref{top13lhc}, we show theoretical results for the double-differential distributions in top-quark transverse-momentum and rapidity, $d^2\sigma/(dp_T dY)$, at 13 TeV LHC energy as functions of the top-quark $p_T$ for given values of the top-quark rapidity in four plots, as before. Again, in each plot we show results at LO, NLO, aNNLO, and aN$^3$LO with a central scale choice of $\mu=m_T$, and scale variation at each order is indicated by the dotted curves. The overall distributions are of course much higher than at 8 TeV but, again, we observe a reduction in scale variation at higher orders. The inset plots with $K$ factors show that the higher-order corrections are important, as is also the case at 8 TeV.

\begin{figure}[htbp]
\begin{center}
\includegraphics[width=83mm]{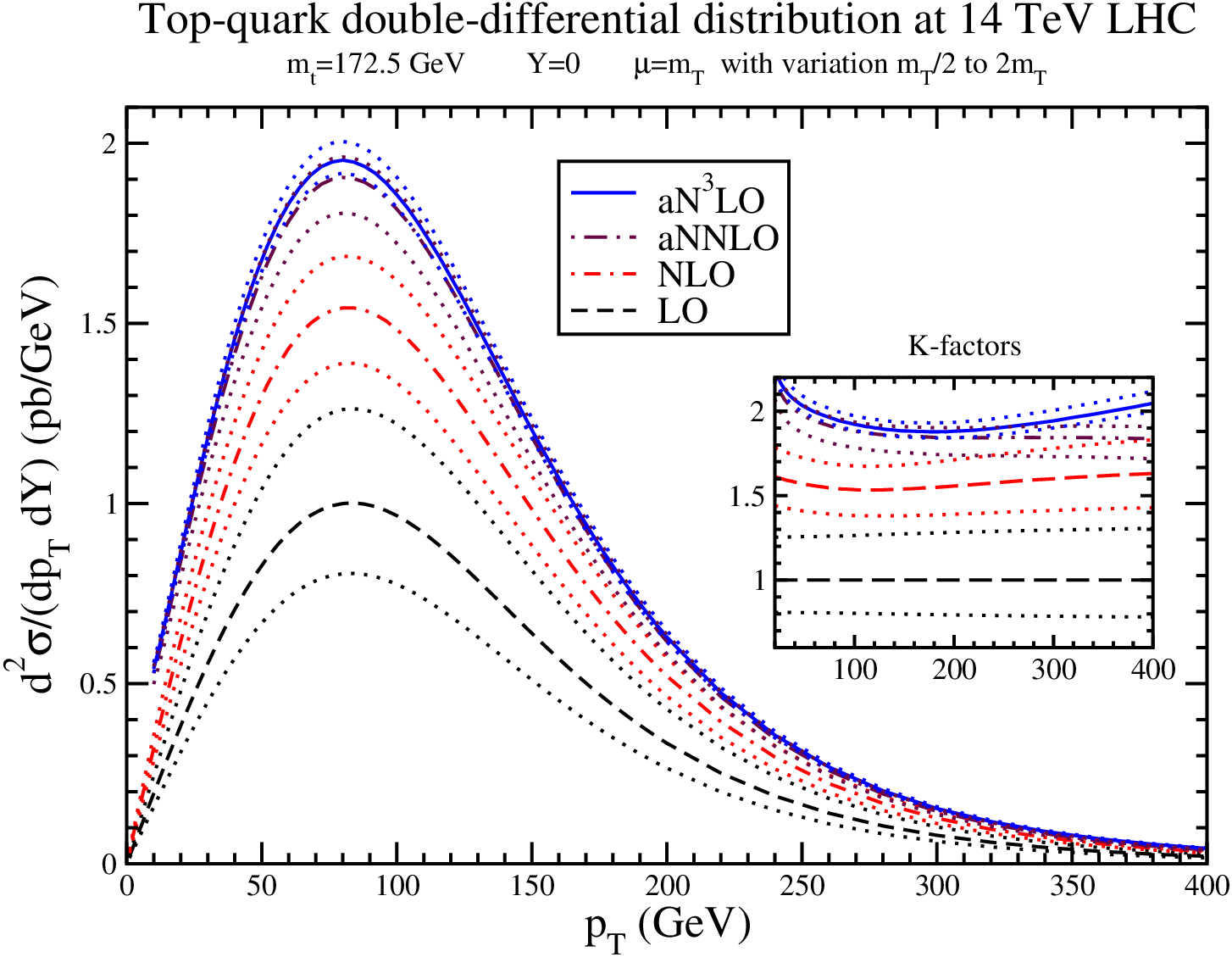}
\hspace{1mm}
\includegraphics[width=83mm]{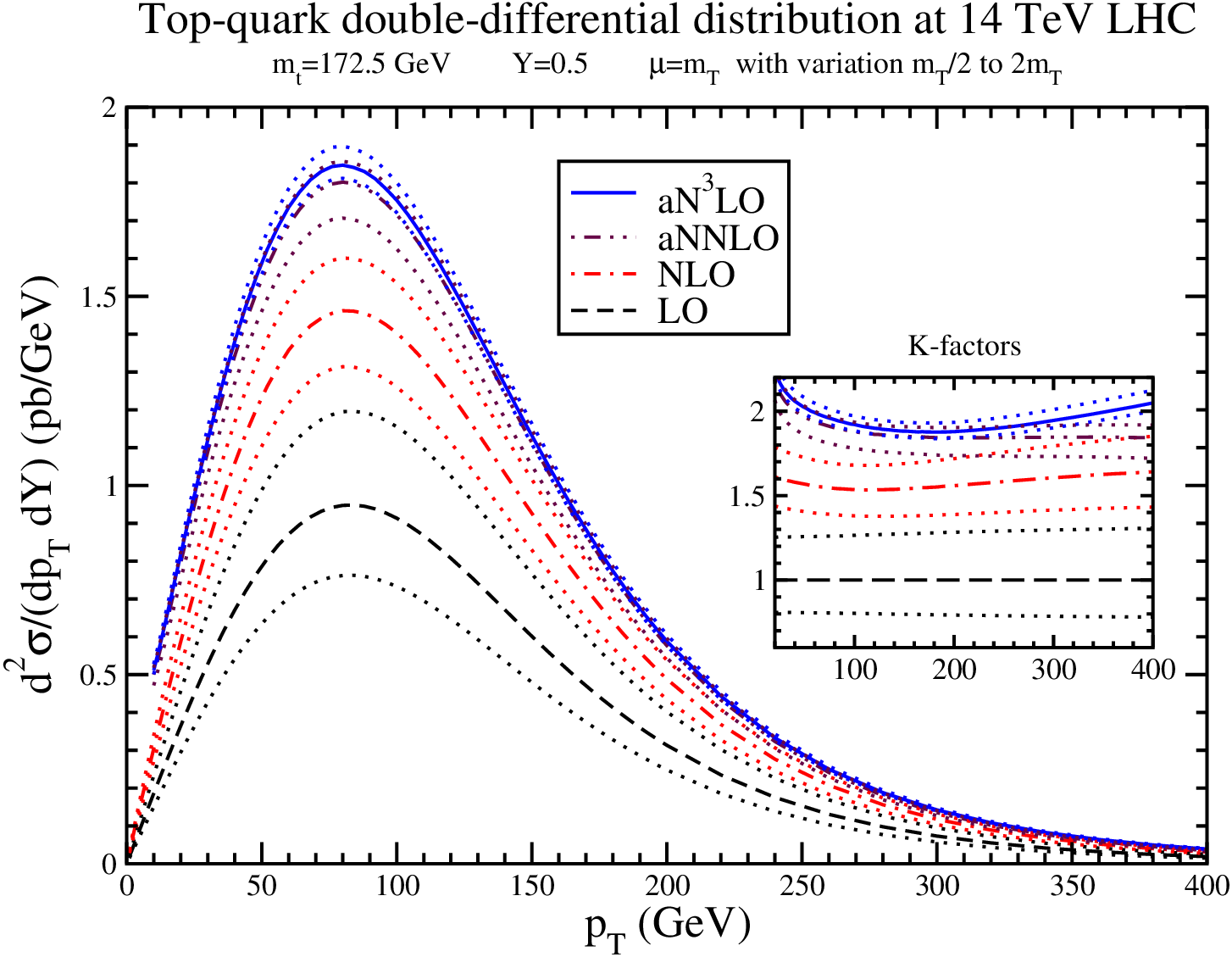}
\bigbreak
\includegraphics[width=83mm]{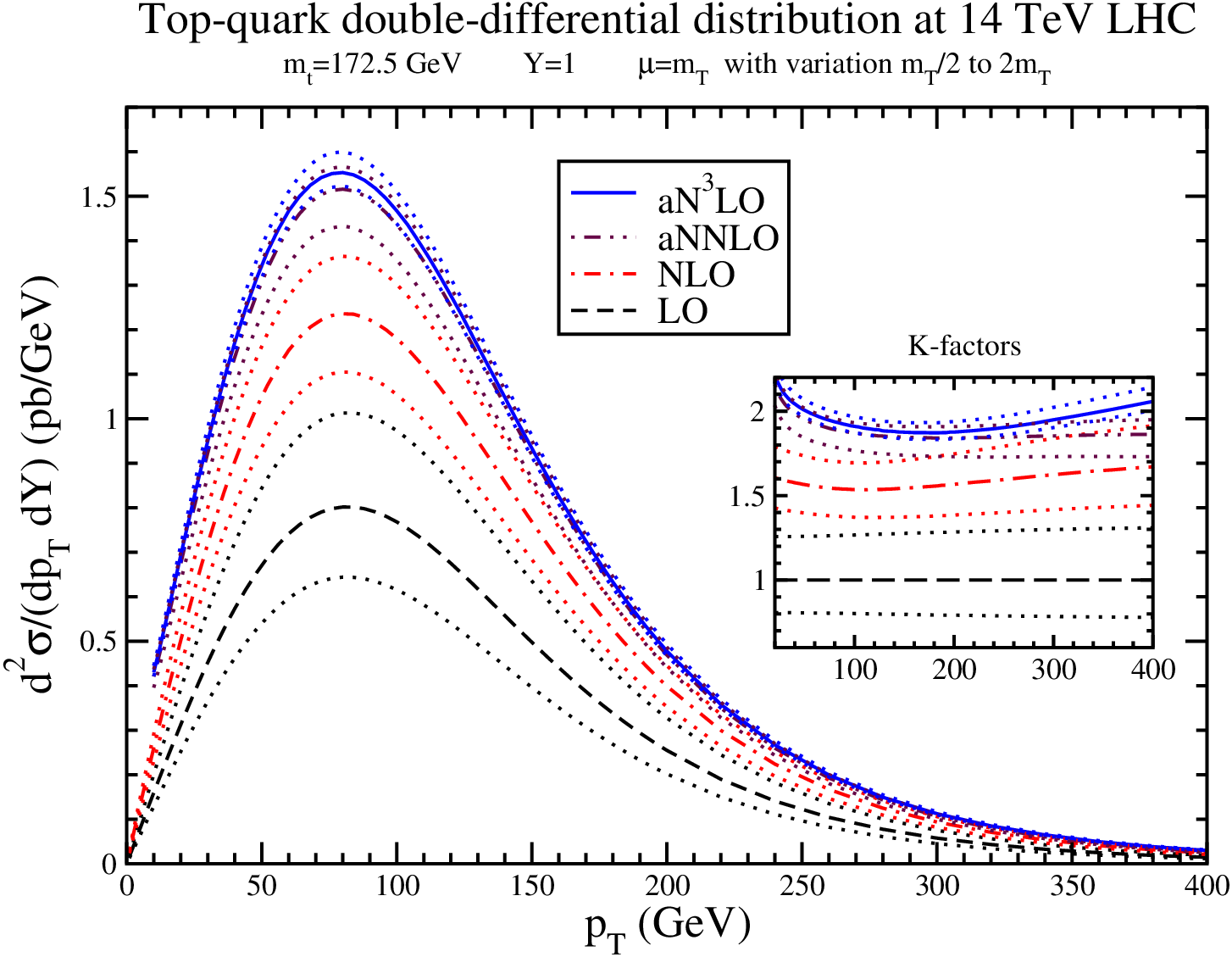}
\hspace{1mm}
\includegraphics[width=83mm]{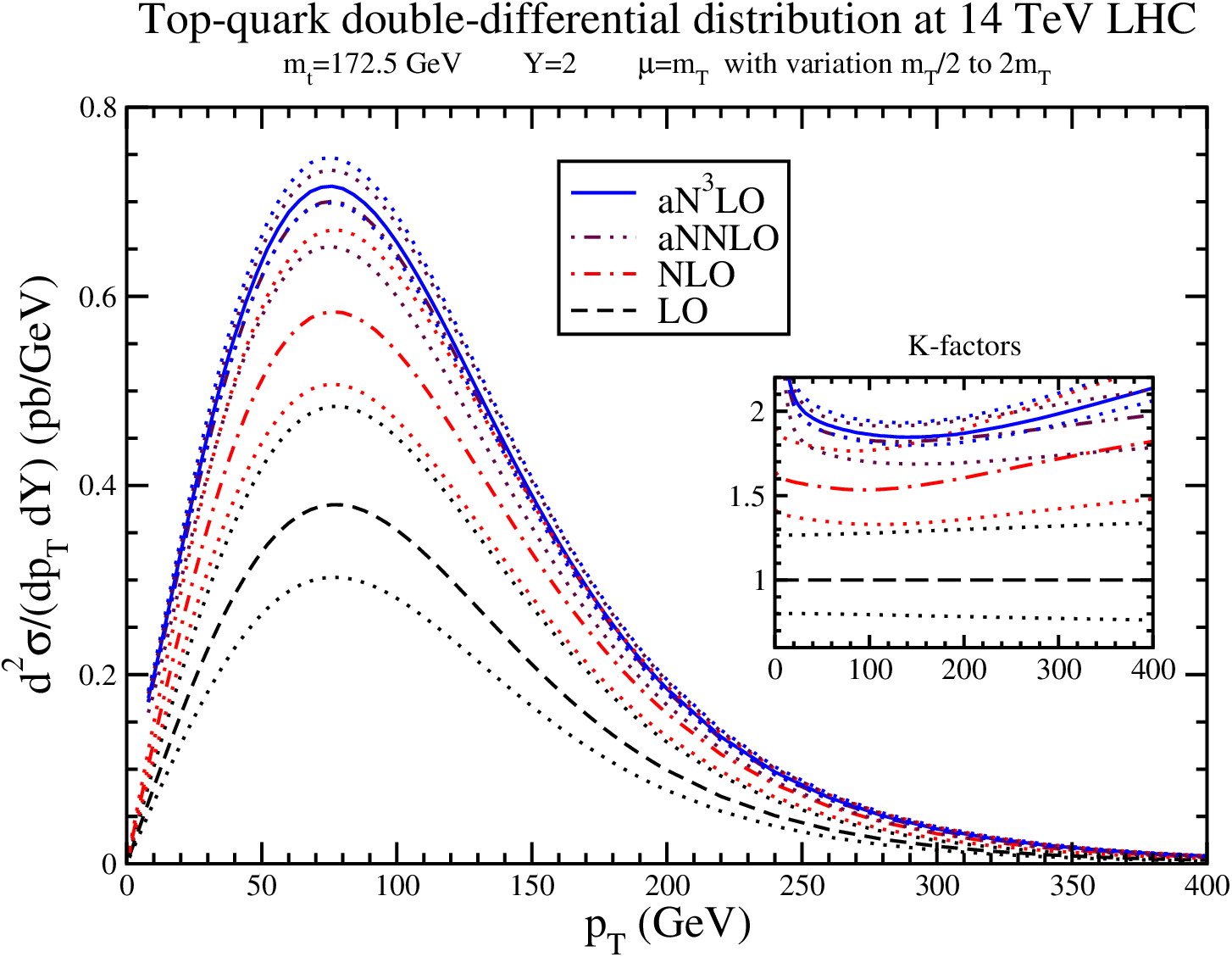}
\caption{The top-quark double-differential distributions, $d^2\sigma/(dp_T dY)$, at 14 TeV LHC energy, are displayed as functions of $p_T$ for four different values of rapidity. The LO, NLO, aNNLO, and aN$^3$LO results are shown with central scale $\mu=m_T$ (solid lines) and scale variation $m_T/2$ and $2m_T$ (dotted lines). The $K$-factors relative to LO are shown in the inset plots.}
\label{top14lhc}
\end{center}
\end{figure}

In Fig. \ref{top14lhc}, we show the corresponding theoretical results for $d^2\sigma/(dp_T dY)$ at 14 TeV LHC energy. Again, central results and scale variation are shown at each order through aN$^3$LO, and, as before, we observe large $K$ factors as well as a reduction in scale variation with increasing perturbative order. The distributions at 14 TeV are somewhat larger than at 13 TeV, and the $K$ factors are very similar.

We next compare our theoretical results with recent data from the CMS experiment at 8 TeV \cite{CMS8tev} and 13 TeV \cite{CMS13tev} LHC energies. Normalized distributions are often used in the presentation of data, since this helps to reduce systematic errors. On the theoretical side, normalized distributions reduce the dependence on the choice of pdf sets. The experimental data for the double-differential distributions are given in terms of discrete transverse-momentum and rapidity bins, in contrast to the smooth theoretical functions that we provided. Therefore, we recalculate the theoretical predictions for the specific binnings used by CMS at 8 and 13 TeV energies. 

\begin{figure}[htbp]
\begin{center}
\includegraphics[width=83mm]{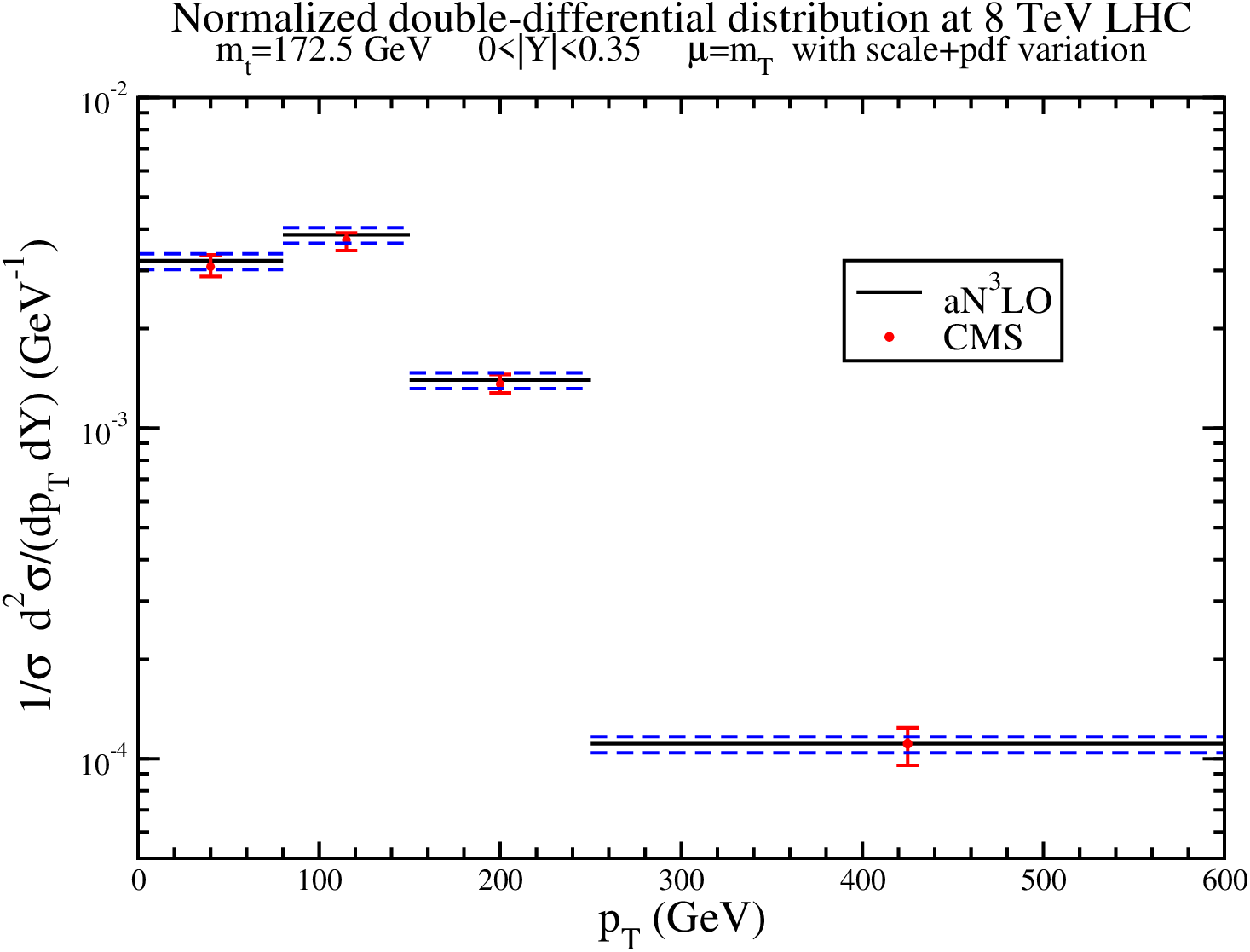}
\hspace{1mm}
\includegraphics[width=83mm]{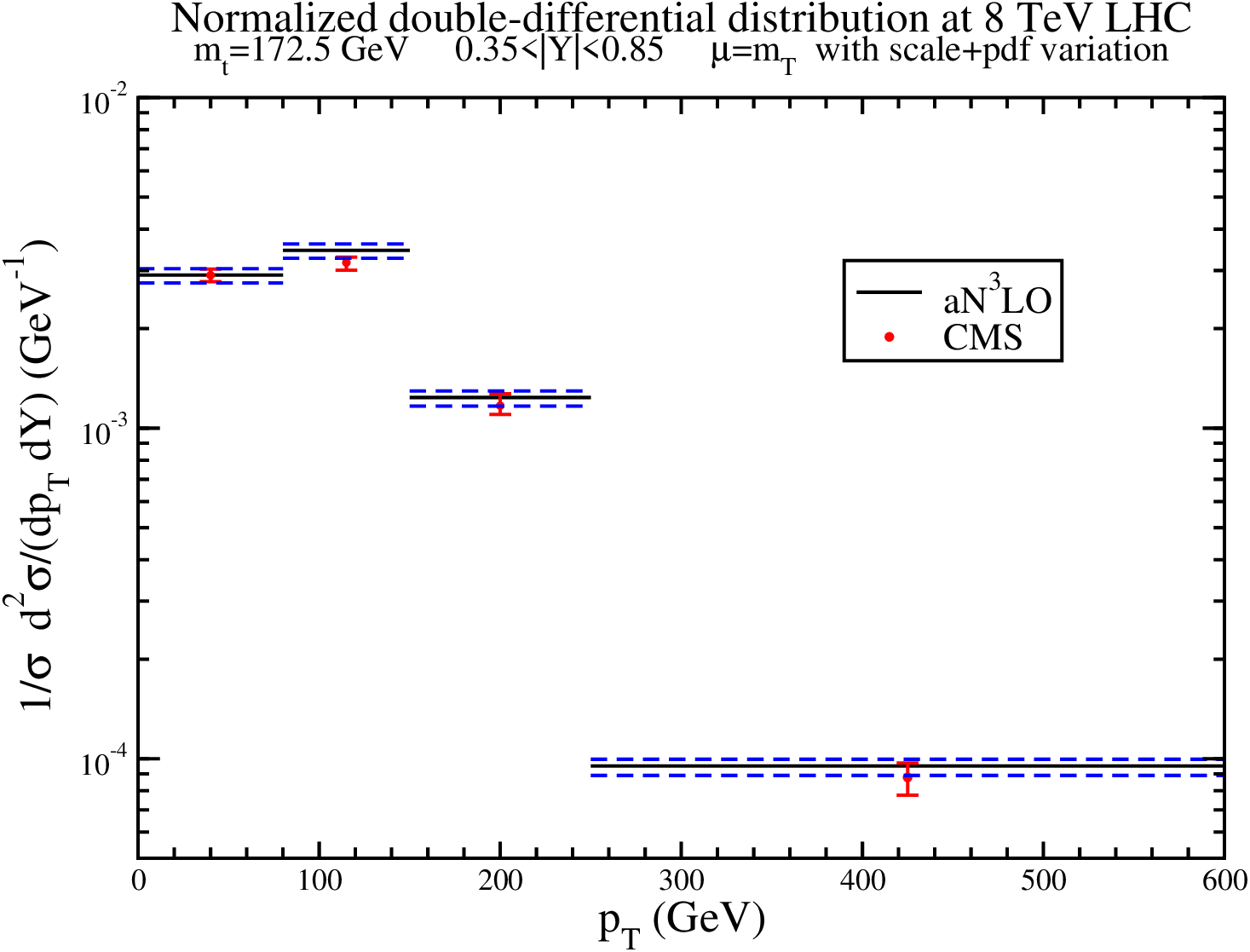}
\bigbreak
\includegraphics[width=83mm]{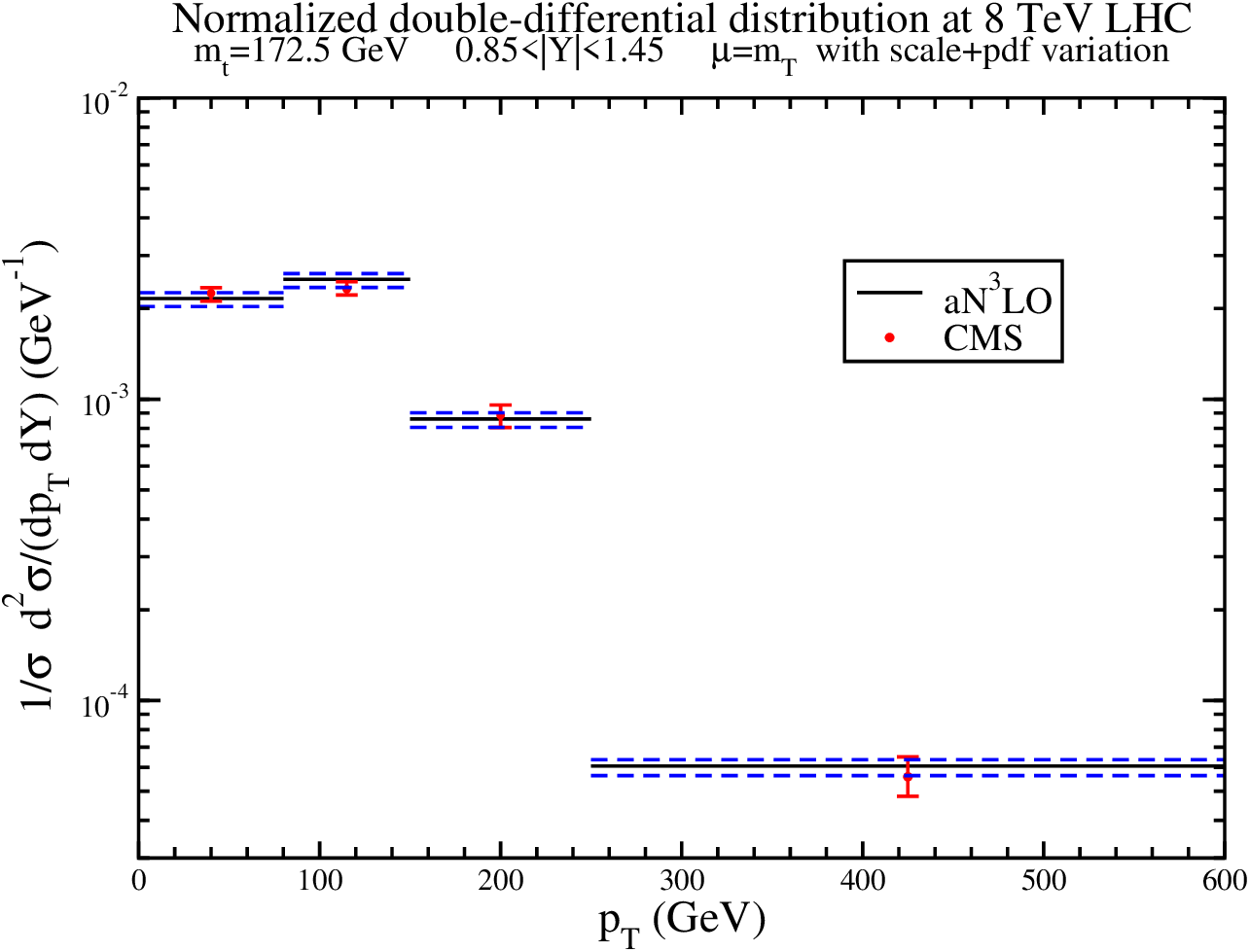}
\hspace{1mm}
\includegraphics[width=83mm]{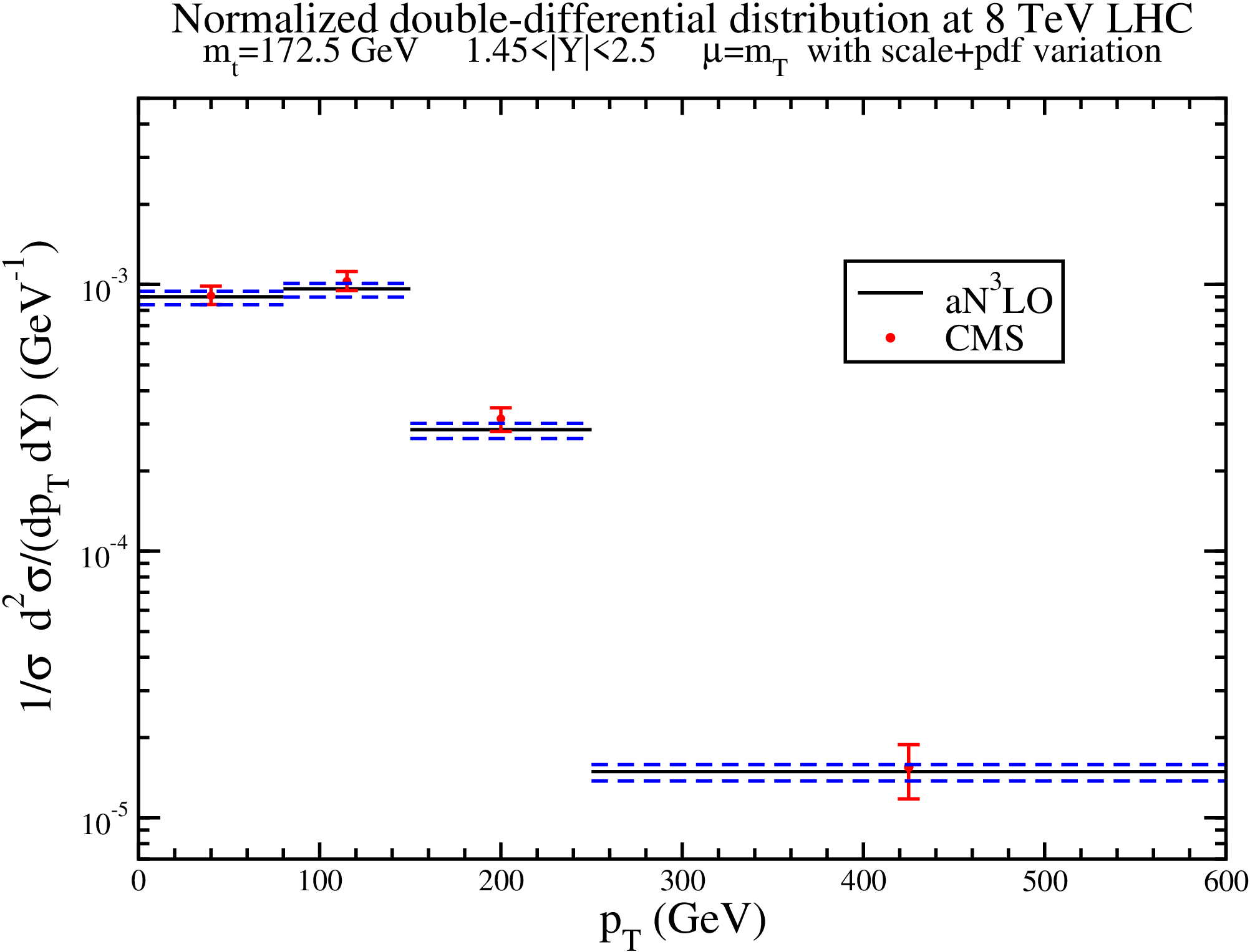}
\caption{The normalized top-quark double-differential distributions, $(1/\sigma) d^2\sigma/(dp_T dY)$, for four different rapidity intervals at 8 TeV LHC energy, are shown at aN$^3$LO (solid lines for $\mu=m_T$) with total theoretical uncertainties from scale variation and pdf uncertainties (dashed lines), and they and compared with CMS data \cite{CMS8tev}.}
\label{diff8lhc}
\end{center}
\end{figure}

In Fig. \ref{diff8lhc} we show the normalized aN$^3$LO double-differential distributions, $(1/\sigma) d^2\sigma/(dp_T dY)$, at 8 TeV LHC energy integrated over and averaged in different $p_T$ bins and rapidity interval $0<|Y|<0.35$ (upper-left plot), $0.35<|Y|<0.85$ (upper-right plot), $0.85<|Y|<1.45$ (bottom-left plot), and $1.45<|Y|<2.5$ (bottom-right plot),  corresponding to the binning from CMS data. We observe very good agreement of theory with data in all four plots. The total theoretical uncertainties, from the combination of scale variation and pdf uncertainties, have also been calculated, and they are indicated in the plots by the dashed curves. We note that the theoretical uncertainties are comparable to the experimental ones in the first three $p_T$ bins, but they are significantly smaller than the experimental ones in the largest $p_T$ bin.

\begin{figure}[htbp]
\begin{center}
\includegraphics[width=83mm]{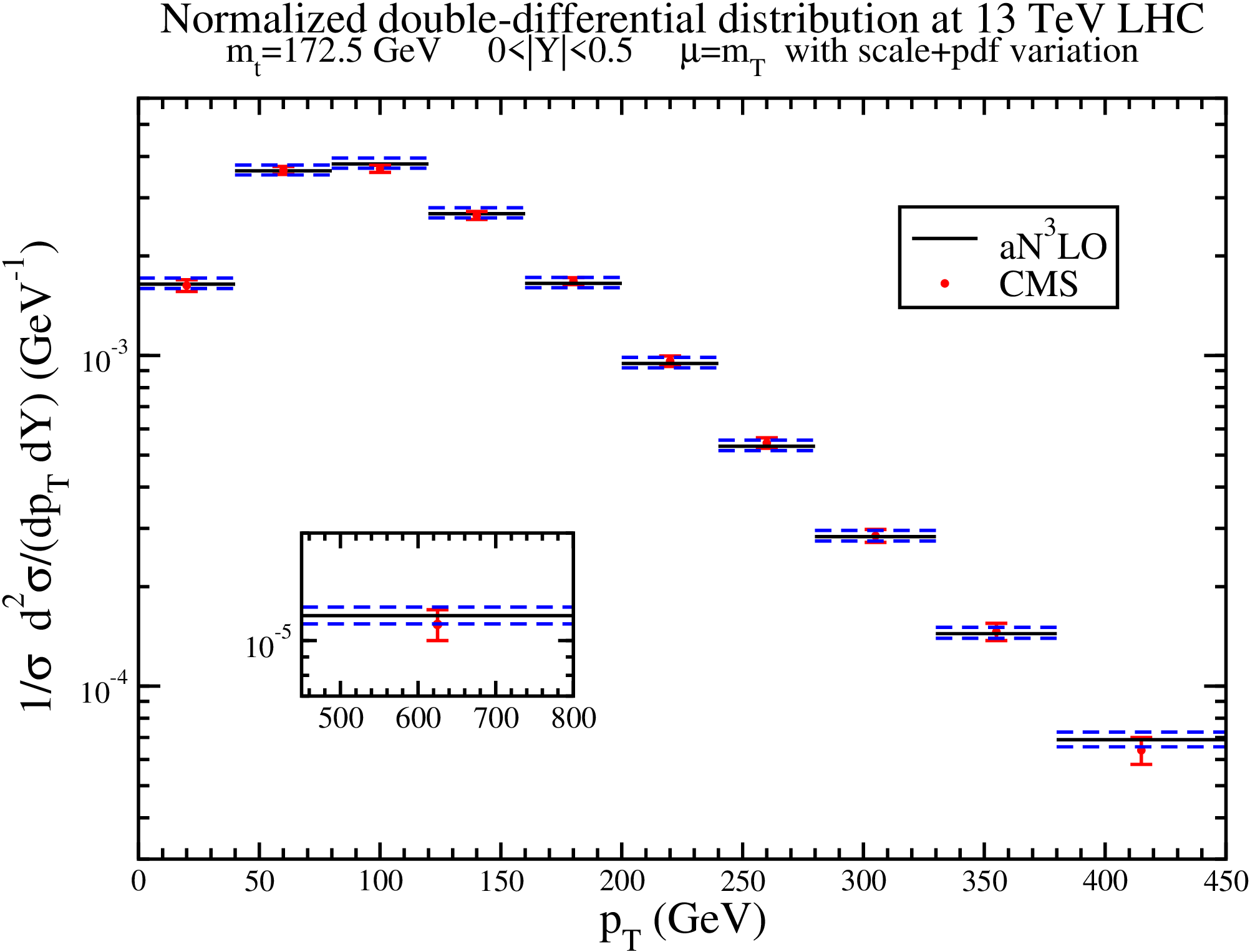}
\hspace{1mm}
\includegraphics[width=83mm]{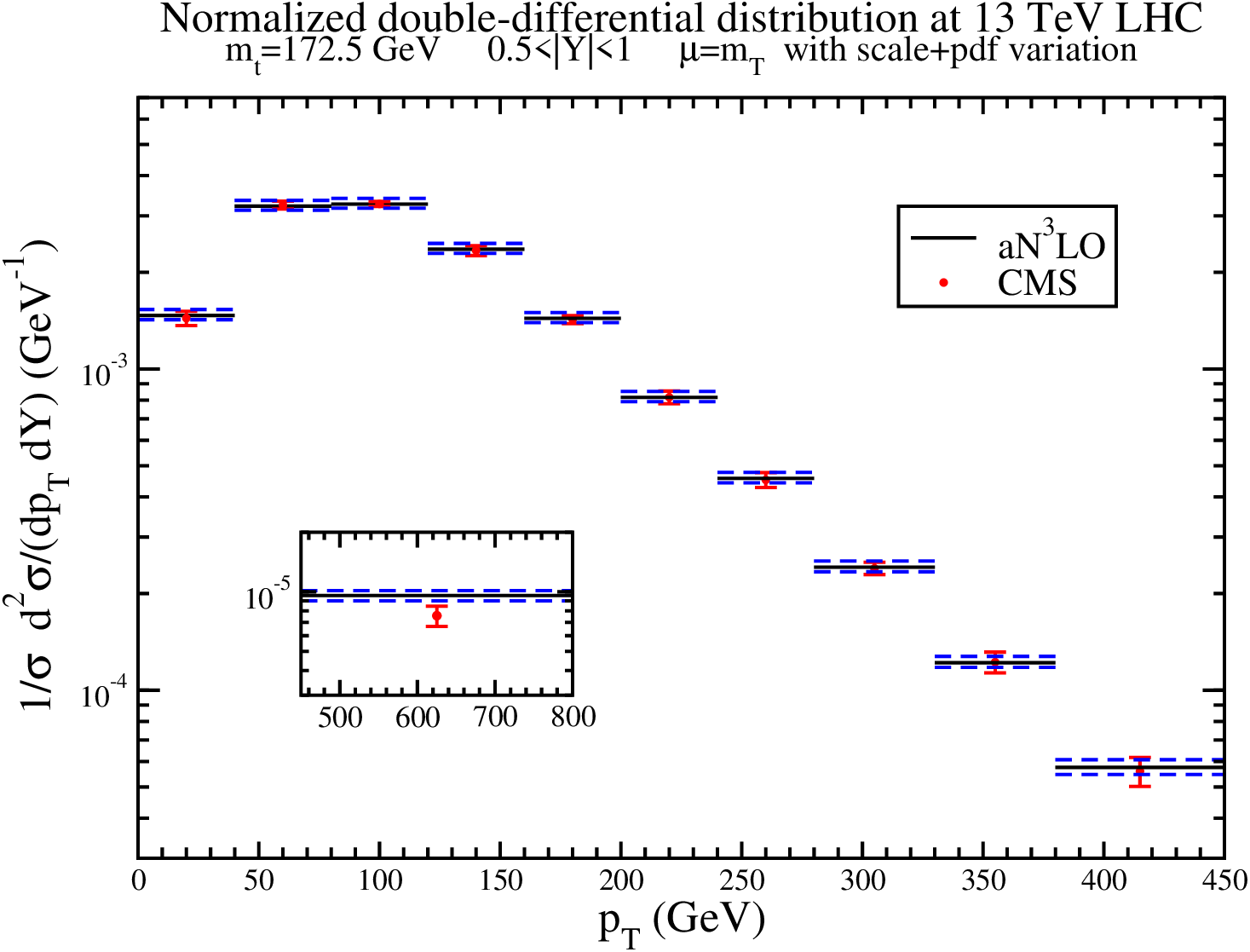}
\bigbreak
\includegraphics[width=83mm]{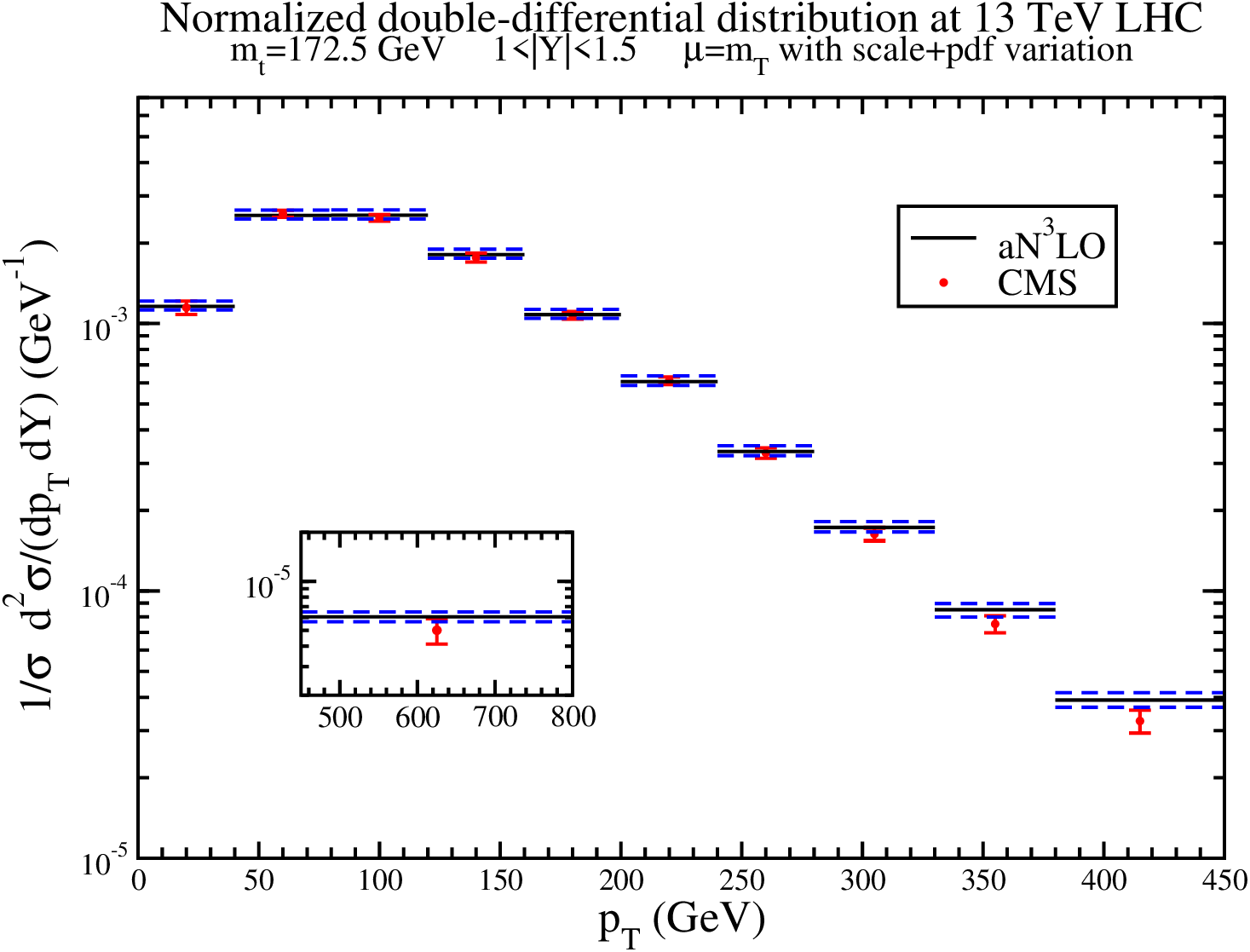}
\hspace{1mm}
\includegraphics[width=83mm]{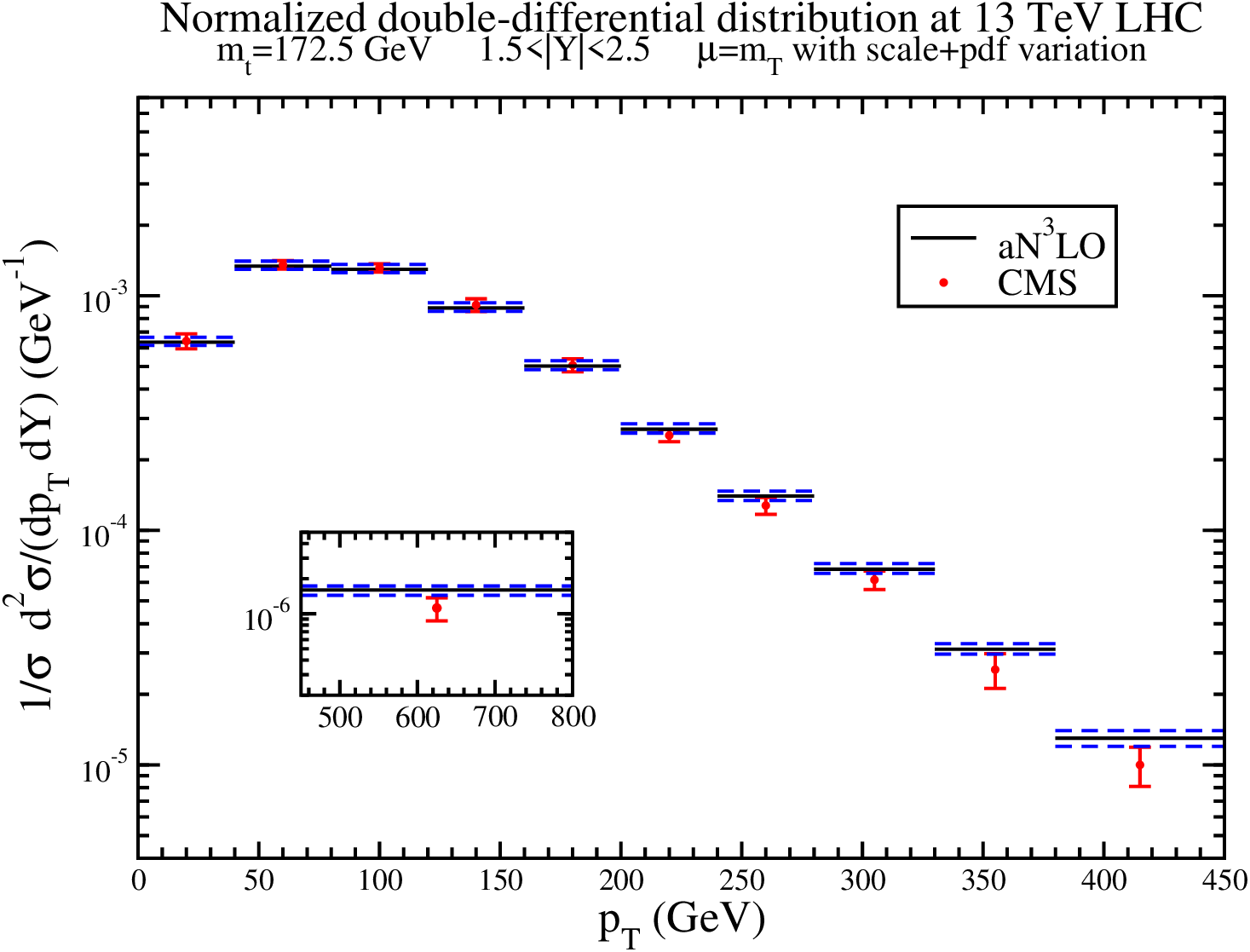}
\caption{The normalized top-quark double-differential distributions, $(1/\sigma) d^2\sigma/(dp_T dY)$, for four different rapidity intervals at 13 TeV LHC energy, are shown at aN$^3$LO (solid lines for $\mu=m_T$) with total theoretical uncertainties from scale variation and pdf uncertainties (dashed lines), and they are compared with CMS data \cite{CMS13tev}.}
\label{diff13lhc}
\end{center}
\end{figure}

In Fig. \ref{diff13lhc} we show the normalized aN$^3$LO double-differential distributions, $(1/\sigma) d^2\sigma/(dp_T dY)$, at 13 TeV LHC energy integrated over and averaged in different $p_T$ bins and rapidity interval $0<|Y|<0.5$ (upper-left plot), $0.5<|Y|<1$ (upper-right plot), $1<|Y|<1.5$ (bottom-left plot), and $1.5<|Y|<2.5$ (bottom-right plot), corresponding to the binning from CMS data. The inset plots display the highest-$p_T$ bins. We observe good agreement of theory with data in all four plots. Again, we calculate the total theoretical uncertainties from scale variation and pdf uncertainties, and note that they are similar to the experimental ones for small and medium $p_T$ values, and significantly smaller than the experimental ones for the larger $p_T$ bins.

\section{Conclusion}

I have presented top-quark double-differential distributions in transverse momentum and rapidity, including soft-gluon corrections through aN$^3$LO. The soft-gluon corrections are derived from NNLL resummation of the double-differential cross section. I have provided detailed theoretical results at 8, 13, and 14 TeV LHC energies. The corrections are important, they substantially increase the rates, and they decrease the scale uncertainties. 

I have also presented predictions for normalized double-differential distributions for specific $p_T$ and rapidity bins used by the CMS experiment at the LHC.
A comparison with CMS data shows that the theoretical results provide a very good description of the data for a wide variety of $p_T$ and rapidity values, at both 8 and 13 TeV LHC energies. Including scale and pdf uncertainties, we observe that the total theoretical uncertainties are either comparable or smaller than the experimental ones.

\section*{Acknowledgements}
This material is based upon work supported by the National Science Foundation under Grant No. PHY 1820795.

\end{document}